\documentclass[preprint,nofootinbib,aps,superscriptaddress]{revtex4}
\usepackage{color,graphicx,shortvrb,epsfig}
\usepackage{subfigure}
\usepackage{amsmath}
\usepackage{amssymb}

\usepackage{color,shortvrb,epsfig}

\usepackage{graphicx,hyperref}
\newcommand{\beq}{\begin{equation}}
\newcommand{\eeq}{\end{equation}}
\newcommand{\beqa}{\begin{eqnarray}}
\newcommand{\eeqa}{\end{eqnarray}}
\newcommand{\no}{\nonumber}
\def\OMIT#1{{}}
\newcommand{\lsim}{\mathrel{\rlap{\lower4pt\hbox{\hskip1pt$\sim$}}
    \raise1pt\hbox{$<$}}}         
\newcommand{\gsim}{\mathrel{\rlap{\lower4pt\hbox{\hskip1pt$\sim$}}
    \raise1pt\hbox{$>$}}}         

\newcommand{\ve}{\varepsilon}
\begin{document}

\preprint{\vbox{\hbox{}\hbox{}\hbox{}
    \hbox{CERN-PH-TH/2005-023}
    \hbox{LMU-ASC 22/05}
    \hbox{WIS/04/05-Mar-DPP}
    \hbox{hep-ph/0503151}}}

\vspace*{1.5cm}

\title{\boldmath The Pattern of CP Asymmetries
  in $b\to s$ Transitions }

\author{Gerhard Buchalla}\email{buchalla@theorie.physik.uni-muenchen.de}
\affiliation{Arnold Sommerfeld Center, Department f\"ur Physik,
  Ludwig-Maximilians-Universit\"at M\"unchen,  
  Theresienstra\ss{}e 37, D-80333 M\"unchen, Germany}

\author{Gudrun Hiller}\email{hiller@theorie.physik.uni-muenchen.de}
\affiliation{Arnold Sommerfeld Center, Department f\"ur Physik,
  Ludwig-Maximilians-Universit\"at M\"unchen, 
  Theresienstra\ss{}e 37, D-80333 M\"unchen, Germany}
\affiliation{CERN, TH Division, Department of Physics,
CH-1211 Geneva 23, Switzerland}

\author{Yosef Nir}\email{yosef.nir@weizmann.ac.il}
\affiliation{Department of Particle Physics,
  Weizmann Institute of Science, Rehovot 76100, Israel}

\author{Guy Raz}\email{guy.raz@weizmann.ac.il}
\affiliation{Department of Particle Physics,
  Weizmann Institute of Science, Rehovot 76100, Israel}


\vspace{2cm}
\begin{abstract}
New CP violating physics in $b\to s$ transitions will modify the CP
asymmetries in $B$ decays into final CP eigenstates ($\phi K_S$,
$\eta^\prime K_S$, $\pi^0 K_S$, $\omega K_S$, $\rho^0 K_S$ and $\eta
K_S$) from their Standard Model values. In a model independent
analysis, the pattern of deviations can be used to probe which Wilson
coefficients get a significant contribution from the new physics. We
demonstrate this idea using several well-motivated models of new
physics, and apply it to current data.  
\end{abstract}

\maketitle

\section{Introduction}
\label{sec:introduction}
CP asymmetries in the decays of neutral $B$ mesons into final CP
eigenstates $f$ exhibit a time-dependent behavior (for a review of CP
violation in meson decays, see \cite{Eidelman:2004wy}),
\beq
{\cal A}_f(t)=S_f\sin(\Delta mt)-C_f\cos(\Delta mt).
\eeq
The Standard Model predicts that for most of the decays that proceed 
via the quark transitions $\bar b\to\bar q^\prime q^\prime\bar s$
($q^\prime=c,s,d,u$), the following relations hold to a good
approximation: 
\beq
-\eta_f S_f\simeq\sin2\beta,\ \ \ \ C_f\simeq0,
\eeq
where $\eta_f=\pm1$ is the CP eigenvalue for the final state $f$, and
$\beta\equiv\phi_1\equiv\arg\left[-(V_{cd}V_{cb}^*)/(V_{td}V_{tb}^*)\right]$.
New physics effects can appear in two ways. First, new physics in
${\cal H}_{\Delta B=2}$ contributes to the $B^0-\overline{B}^0$ mixing
amplitude. Such a contribution shifts all $S_f$'s in a universal way:
The $S_f$ asymmetries remain equal to each other, though
different from $\sin2\beta$. The $C_f$'s still
vanish. Second, new physics in ${\cal H}_{\Delta B=1}$ can contribute
to the $\bar b\to\bar qq\bar s$ transitions ($q=s,d,u$) \cite{history}. (The tree
level transition $\bar b\to\bar cc\bar s$ is
unlikely to get a significant contribution from new physics, and
consequently the asymmetries $S_{\psi K}$ and $C_{\psi K}$ would not
be modified by ${\cal H}_{\Delta B=1}$.) Such a contribution would
lead to interesting consequences:
\begin{enumerate}
  \item The $S_f$'s would be different from each other and from
    $S_{\psi K}$. 
  \item The $C_f$'s would be different from each other and from zero.
  \end{enumerate}
Thus, in the presence of new physics, a pattern of deviations, $\delta
S_f=-\eta_f S_f-S_{\psi K}\neq0$ and $C_f\neq0$, will arise.

Given a set of experimental ranges for $\delta S_f$ and $C_f$, one
would like to interpret these data in terms of new physics. There are
two ways to proceed. First, one can choose a model of new physics and
analyze whether the model can accommodate the data. The second way is
model independent. The effects of new physics can be described by a
modification of the Wilson coefficients in the operator product
expansion for $\Delta B=1$ interactions. Thus, one can fit a set of
Wilson coefficients to the data and learn which operators can account
for the observed deviations. In this paper, we study this second,
model independent, method.

While our analysis has the advantage of being model independent, it
suffers from two limitations:
\begin{itemize}
  \item The number of Wilson coefficients is larger than the number of
    measured CP asymmetries. Consequently, an analysis that is
    entirely generic is impossible to carry out at present. Hence, one
    has to assume 
    that only a subset of all possible operators are modified. In this
    work we consider some simple cases, where the new physics is
    parameterized by a single complex parameter. The cases that we
    study are motivated by actual models of new physics. Extensions to
    a larger number of new physics parameters is left for future
    work. It is also possible to include in the analysis additional
    data, beyond the CP asymmetries in decays into final CP
    eigenstates. This extension is also left for future work.
    \item To interpret the data in terms of Wilson coefficients, one
      has to know the values of the (mode-dependent) hadronic matrix
      elements of the operators. At present, there is no first
      principle calculation of the matrix elements that has been
      tested to a high level of precision. Thus, hadronic
      uncertainties prevent a clean theoretical interpretation. 
      We use factorization \cite{Beneke:2000ry,Ali:1998eb,Beneke:2002jn} 
      for our
      analysis.\footnote{In this work we perform our calculations to
      leading order in $\alpha_s$ and drop $1/m_b$ corrections, except 
      for the chirally enhanced terms
      related to $a_6$ and $a_8$. In this approximation, QCD- and
      naive-factorization are identical. We note, however,
      differences in the expressions for the $f=\eta^{(\prime)}K$
      modes between refs. \cite{Ali:1998eb} and \cite{Beneke:2002jn}:
      A term $\propto f^u_{\eta^{(\prime)}}(a_6-\frac12 a_8)$ should
      be omitted from the first, while a term $\propto
      f^c_{\eta^{(\prime)}}(a_7-a_9)$ should be added to the second.}   
      \end{itemize}

A number of relevant measurements exists already. The experimental
situation is summarized in Table \ref{tab:fcp}.

\begin{table}[t]
\caption{World averages of the experimental results for the CP
  asymmetries in neutral $B$ decays via the quark transitions $\bar 
  b\to\bar q^\prime q^\prime\bar s$ modes, $q^\prime=c,s,d,u$. \label{tab:fcp}}
\vspace{0.4cm}
\begin{center}
\begin{tabular}{|c|c|c|c|}
  \hline
$f$ & $-\eta_{\rm CP}S_f$ & $C_f$ & Refs.\\ \hline 
$\psi K_S$ & $+0.73\pm0.04$ & $+0.03\pm0.03$ &
\ \cite{Abe:2004mz,Aubert:2004zt}\ \\ \hline
$\phi K_S$ & $+0.35\pm0.21$ & $-0.04\pm0.17$ &
\ \cite{Aubert:2005ja,Abe:2004xp}\ \\
$\eta^\prime K_S$ & $+0.43\pm0.11$ & $-0.04\pm0.08$ &
\ \cite{Aubert:2005iy,Abe:2004xp}\ \\
$\pi^0 K_S$ & $+0.34\pm0.29$ & $+0.08\pm0.14$ &
\ \cite{Aubert:2005zt,Abe:2004xp}\ \\
 $\omega K_S$ & $+0.56\pm0.32$ & $-0.49\pm0.25$ &
\ \cite{Abe:2004xp,Aubert:2005ra}\ \\ \hline
\end{tabular}
\end{center}
\end{table}

The plan of this paper is as follows. In section \ref{sec:for} we
introduce our formalism and evaluate, to leading log approximation and
within the QCD factorization approach, the Standard Model (SM)
predictions for the CP asymmetries. In section \ref{sec:np} we analyze
the effects of new physics. We start with a generic, model independent
analysis, using the operator product expansion for $\Delta B=1$
operators. We then focus on scenarios where the effects of new physics
depend on a single complex parameter. We give three explicit examples
of such scenarios. In section \ref{sec:pre} we apply our approach to
present data. We conclude in section \ref{sec:con}.

\section {Formalism and Standard Model Predictions}
\label{sec:for}
We follow the notations of ref. \cite{Beneke:2000ry}. We consider the
following  effective Hamiltonian for $\Delta B=\pm1$ decays:
\beq\label{heffbone}
{\cal H}_{\rm eff}={G_F\over\sqrt2}\sum_{p=u,c}
  V_{ps}^*V_{pb}\left(C_1O_1^p+C_2O_2^p+ 
  \sum_{i=3}^{10}C_iO_i+C_{7\gamma}O_{7\gamma}+C_{8g}O_{8g}\right)+{\rm h.c.},
\eeq
with
\beqa\label{curcur}
O_1^p=(\bar pb)_{V-A}(\bar sp)_{V-A},&\ \ \ \ \ &
O_2^p=(\bar p_{\beta}b_{\alpha})_{V-A}(\bar s_{\alpha}p_{\beta})_{V-A},\no\\
O_3=(\bar sb)_{V-A}\sum_{q}(\bar qq)_{V-A},&\ \ \ \ \ & 
O_4=(\bar s_\alpha b_{\beta})_{V-A}\sum_{q}(\bar q_{\beta}q_{\alpha})_{V-A},\nonumber\\
O_5=(\bar sb)_{V-A}\sum_{q}(\bar qq)_{V+A},&\ \ \ \ \ &
O_6=(\bar s_{\alpha}b_{\beta})_{V-A}\sum_{q}(\bar q_{\beta}q_{\alpha})_{V+A},\no\\
O_7=\frac{3}{2}(\bar sb)_{V-A}\sum_{q}e_q(\bar qq)_{V+A},&\ \ \ \ \ &
O_8=\frac{3}{2}(\bar s_{\alpha}b_{\beta})_{V-A}\sum_{q}e_q(\bar
q_{\beta}q_{\alpha})_{V+A},\nonumber\\
O_9=\frac{3}{2}(\bar sb)_{V-A}\sum_{q}e_q(\bar qq)_{V-A},&\ \ \ \ \ &
O_{10}=\frac{3}{2}(\bar s_{\alpha}b_{\beta})_{V-A}\sum_{q}e_q(\bar q_{\beta}q_{\alpha})_{V-A},\no\\
O_{7\gamma}=-\frac{em_b}{8\pi^2}\bar s\sigma^{\mu\nu}(1+\gamma_5) F_{\mu\nu}b,&\ \ \ \ \ & 
O_{8g}=-\frac{g_sm_b}{8\pi^2}\bar s\sigma^{\mu\nu}(1+\gamma_5)G_{\mu\nu}b,
\eeqa
where $(\bar q_1 q_2)_{V\pm A}=\bar q_1\gamma_\mu(1\pm\gamma_5)q_2$,
the sum is over active quarks, with $e_q$ denoting their electric
charge in fractions of $|e|$ and $\alpha,\beta$ are color indices. 

The CP asymmetries in $B\to f$ decays are calculated as follows. One
defines a complex quantity $\lambda_f$,
\beq
\lambda_f=e^{-i\phi_B}(\overline{A}_f/A_f),
\eeq
where $\phi_B$ is the phase of $M_{12}$, the $B^0-\overline{B}^0$
mixing amplitude, and $A_f$ ($\overline{A}_f$) is the decay amplitude
for $B^0(\overline{B}^0)\to f$. We have
\beq
S_f=\frac{2{\cal I}m(\lambda_f)}{1+|\lambda_f|^2},\ \ \
C_f=\frac{1-|\lambda_f|^2}{1+|\lambda_f|^2}.
\eeq
The decay amplitudes can be calculated from the effective Hamiltonian
of eq. (\ref{heffbone}) \cite{Ali:1998eb}:
\beq
A_f=\langle f|{\cal H}_{\rm eff}|B^0\rangle,\ \ \
\overline{A}_f=\langle f|{\cal H}_{\rm eff}|\overline{B}^0\rangle.
\eeq
The electroweak model determines the Wilson coefficients    
while QCD (or, more practically, a calculational method such as QCD
factorization) determines the matrix elements $\langle
f|O_i|B^0(\overline{B}^0)\rangle$.

We perform our calculations to leading-log approximation. In 
particular, to run the Wilson coefficients from the weak scale $m_W$
to a low scale of order $m_b$, we use the 12-dimensional leading-log
anomalous dimension matrix $\gamma$ \cite{Buchalla:1995vs} that is
given in Table \ref{AD}. The $\alpha_s$ mixing of electroweak penguins
onto the dipole operators is deduced from
\cite{Buchalla:1995vs,Chetyrkin:1997gb,Baranowski:1999tq}. 
\begin{table}
\caption{The anomalous dimension matrix at leading log. The rows and
  columns correspond to the twelve operators
  $O_{1,\ldots,10,7\gamma,8g}$ defined in eq. (\ref{curcur}).}
\vspace{0.4cm}
 \begin{center}
\begin{tabular}{|cc|cccc|cccc|cc|}
\hline
$-2$ & 6 & $ -\frac{2}{9} $ &  $ \frac{2}{3} $ &  $-\frac{2}{9} $ &
$\frac{2}{3}$
& $ 0 $ & $ 0 $ & $ 0 $ & $ 0 $ & $ \frac{416}{81} $ & $ \frac{70}{27} $
\\
6 & $-2$ & 0 & 0 & 0 & 0 & 0 & 0 & 0 & 0 & 0 & 3 \\
\hline
$0 $ & $ 0 $ & $ -\frac{22}{9} $ & $ \frac{22}{3} $ & $ -\frac{4}{9} $ &
$ \frac{4}{3} $ & $
 0 $ & $ 0 $ & $ 0 $ & $ 0 $ & $ -\frac{464}{81} $ & $ \frac{545}{27} $ \\
$ 0 $ & $ 0 $ & $ \frac{44}{9} $ & $ \frac{4}{3} $ & $ -\frac{10}{9} $ &
$ \frac{10}{3} $ & $
 0 $ & $ 0 $ & $ 0 $ & $ 0 $ & $ \frac{136}{81} $ & $ \frac{512}{27} $ \\
$0 $ & $ 0 $ & $ 0 $ & $ 0 $ & $ 2 $ & $ -6 $ & $
 0 $ & $ 0 $ & $ 0 $ & $ 0 $ & $ \frac{32}{9} $ & $ -\frac{59}{3}$ \\
$0 $ & $ 0 $ & $ -\frac{10}{9} $ & $ \frac{10}{3} $ & $ -\frac{10}{9} $ &
$ -\frac{38}{3} $ & $
 0 $ & $ 0 $ & $ 0 $ & $ 0 $ & $ -\frac{296}{81} $ & $ -\frac{703}{27} $
\\
\hline
$ 0 $ & $ 0 $ & $ 0 $ & $ 0 $ & $ 0 $ & $ 0 $ & $ 2 $ & $ -6 $ & $ 0 $ & $
0 $& $ -\frac{16}{9} $ & $ \frac{5}{6} $ \\
$ 0 $ & $ 0 $ & $ -\frac{1}{9} $ & $ \frac{1}{3} $ & $ -\frac{1}{9} $ &
$ \frac{1}{3} $
& $ 0 $ & $ -16 $ & $ 0 $ & $ 0 $ & $ -\frac{1196}{81} $ &
$ -\frac{11}{54}$  \\
$0 $ & $ 0 $&$ \frac{2}{9} $&$ -\frac{2}{3} $&$ \frac{2}{9} $&$
-\frac{2}{3}
$&$ 0 $&$ 0 $&$ -2 $&$ 6 $&$ \frac{232}{81} $&$ -\frac{59}{54} $ \\
$0 $&$ 0 $&$ -\frac{1}{9} $&$ \frac{1}{3} $&$ -\frac{1}{9} $&$ \frac{1}{3}
$&$ 0 $&$ 0 $&$ 6 $&$ -2 $&$ \frac{1180}{81} $&$ -\frac{46}{27} $ \\
\hline
$0 $&$ 0 $&$ 0 $&$ 0 $&$ 0 $&$ 0 $&$ 0 $&$ 0 $&$ 0 $&$ 0 $&$ \frac{32}{3}
$
&$ 0 $ \\
$0 $&$ 0 $&$ 0 $&$ 0 $&$ 0 $&$ 0 $&$ 0 $&$ 0 $&$ 0 $&$ 0 $
&$ -\frac{32}{9} $&$ \frac{28}{3} $ \\
\hline
\end{tabular}
\label{AD}
\end{center}
\end{table}

Within the SM, we have the following set of Wilson coefficients at
leading order: 
\beq\label{cimwsm}
C_1^{\rm SM}(m_W)=1,\ \ \ C_{i\neq1}^{\rm SM}(m_W)=0.
\eeq
(Strictly speaking, $C_{7\gamma}^{\rm SM}(m_W)$ and $C_{8g}^{\rm
  SM}(m_W)$ are also different from zero. However, their contributions
to the decay processes of interest occur at next-to-leading order
which we neglect in the present work.) 
The SM contribution to the decay amplitudes, related to $\bar b\to
\bar q^\prime q^\prime \bar s$ transitions, can always be written as a 
sum of two terms, $A_f^{\rm SM}=A_f^c+A_f^u$, with
$A_f^c\propto V_{cb}^*V_{cs}$ and $A_f^u\propto
V_{ub}^*V_{us}$. Defining the ratio $a_f^u\equiv
e^{-i\gamma}(A_f^u/A_f^c)$, where
$\gamma\equiv\phi_3\equiv\arg\left[-(V_{ud}V_{ub}^*)/(V_{cd}V_{cb}^*)\right]$,  
we have   
\beq\label{defafu}
A_f^{\rm SM}=A_f^c\left(1+a_f^u e^{i\gamma}\right).
\eeq
For $B\to\psi K_S$, the $a_f^u$ term can be safely neglected (its
effects are below the percent level) and, consequently,
\beq
\lambda_{\psi K_S}^{\rm SM}=-e^{-2i\beta}.
\eeq
For charmless modes, the effects of the $a_f^u$ terms (often called
`the SM pollution') are at least of order
$|(V_{ub}V_{us}^*)/(V_{cb}V_{cs}^*)|\sim$ a few percent. They can lead
to a deviation of $S_f$ from $S_{\psi K}$ and of $C_f$ from zero. In
Table \ref{tab:afu} we give the values of the $a_f^u$ parameter
(obtained, as explained above, by using factorization
\cite{Beneke:2000ry,Ali:1998eb,Beneke:2002jn}) for
all relevant modes. Since we perform our calculations to leading log
approximation, we neglect, for consistency, also power corrections
within the QCD factorization approach, except for the dominant
chirally-enhanced ones related to
    $a_6$ and $a_8$. This means that the strong phases
vanish, and that $C_f=0$ for all $f$ in this approximation. The SM
predictions for $S_f$ (with $S_{\psi K_S}=0.73$ and $\gamma=62^o$
\cite{Charles:2004jd} taken as input) are  also given in Table \ref{tab:afu}.
The values of the input parameters used are specified in Table \ref{tab:input}.
To estimate our uncertainties, we give the values
for $\mu=m_b=4.2$ GeV and for a varied
renormalization scale of $\mu= m_b/2$ and  $\mu=2 m_b$.
Note that  the coefficients of the chirally enchanced terms, defined as 
$r_\chi=2 m_K^2/(m_b m_s)$
and $r_\chi^{S_s}=h_S^s/(f_S^s m_b m_s)$ for the 
singlets $S=\eta, \eta^\prime$ (see \cite{Beneke:2002jn} for details),
evolve with the renormalization scale through the running 
($\overline{\rm MS}$-bar) quark masses.


\begin{table}[t]
  \caption{The $a_f^u$ parameters, calculated in QCD factorization at
    leading log and to zeroth order in $\Lambda/m_b$
    (except for chirally enhanced corrections), and 
    the SM values of $S_f$ for $\mu=m_b$ and in parentheses the respective values for $\mu=2 m_b$ (first) and $\mu= m_b/2$ (second) if different from the central one. \label{tab:afu}}
\vspace{0.4cm}
\begin{center}
  \begin{tabular}{|c|c|c|}
  \hline
$f$               & $a_f^u$  & $-\eta_{\rm CP}S_f$ \\ \hline 
$\psi K_S$        & $0$      & $0.73$  \\ \hline
$\phi K_S$        & $0.019$  & $0.75$  \\
$\pi^0 K_S$       & $0.052 \, \left[0.094, 0.021\right]$  & $0.79 \, \left[0.83, 0.76\right]$  \\
$\eta K_S$        & $0.08 \, \left[0.16, 0.02\right] $   & $0.82 \, \left[0.88, 0.76\right]$  \\
$\eta^\prime K_S$ & $0.007 \, \left[-0.006, 0.019\right] $ & $0.74 \, \left[0.72, 0.75\right]$  \\
$\omega K_S$      & $0.22 \, \left[0.37, 0.04\right]$   & $0.92 \, \left[0.98, 0.78\right]$  \\
$\rho^0 K_S$      & $-0.16 \, \left[-0.32, 0.005\right]$  & $0.49 \, \left[0.19, 0.74\right]$  \\ 
\hline
\end{tabular}
\end{center}
\end{table}

\begin{table}
  \caption{Input parameters used in the calculations.}
\vspace{0.4cm}
  \centering
  \begin{tabular}{|@{\hspace{2em}}c@{\hspace{2em}}|c|c@{\hspace{1em}}|c|c|c@{\hspace{1em}}|c|c|}
    \cline{1-2} \cline{4-5} \cline{7-8}
    \multicolumn{2}{|c|}{Decay constants (MeV)} &&
    \multicolumn{2}{|c|}{Form Factors} &&    \multicolumn{2}{|c|}{$r_\chi$-factors,  quark masses} \\
    \cline{1-2} \cline{4-5} \cline{7-8}
    $f_\pi$ & $131$ && $F^{B \to \pi}(0)$ & $0.28$ &&&  \\
    $f_K$ & $160$ && $F^{B \to K}(0)$ & $0.34$ && $r_\chi(m_b)$ &
    $1.170$ \\
    $f^q_{\eta}$ & $108$ && $F^{B \to \eta}(0)$ & $0.23$ &&
    $r_\chi^{\eta_s}(m_b)$ & $1.230$ \\
    $f^s_{\eta}$ & $-111$ &&&&&& \\
    $f^q_{\eta'}$ & $89$ && $F^{B \to \eta^\prime}(0)$ & $0.19$ &&
    $r_\chi^{\eta'_s}(m_b)$ & $1.241$ \\ \cline{7-8}
    $f^s_{\eta'}$ & $136$ &&&&& $m_b(m_b)$& 4.2 GeV \\
    $f_\omega$ & $187$ && $A^{B \to \omega}(0)$ & $0.33$ && $m_s$(2GeV)& 110 MeV \\
    $f_\rho$ & $209$ && $A^{B \to \rho}(0)$ & $0.37$ && $(m_u+m_d)$(2GeV)& 9.1 MeV \\
    \cline{1-2} \cline{4-5} \cline{7-8}
  \end{tabular}
  \label{tab:input}
\end{table}


An examination of Table \ref{tab:afu} shows that the SM pollution is
small (that is, at the naively expected level of
$|(V_{ub}V_{us}^*)/(V_{cb}V_{cs}^*)|\sim$ a few percent) for $f=\phi K_S,\
\eta^\prime K_S$ and $\pi^0 K_S$. It is larger for $f=\eta
K_S,\ \omega K_S$ and $\rho^0 K_S$. In these modes, $a_f^u$ is
enhanced because, within the QCD factorization approach, there is an
accidental cancellation between the leading contributions to $A_f^c$.
The reason for the suppression of the leading $A^c_f$ piece in 
$f=\rho K,\ \omega K$ versus $f=\pi^0 K$ is that the dominant QCD-penguin
coefficients $a_4$ and $a_6$ appear in $A_{(\rho,\omega)K}^c$ as
$(a_4 -r_\chi a_6)$ and in $A^c_{\pi^0 K}$ as
$(a_4 +r_\chi a_6)$. Since $r_\chi \simeq 1$ and,
within the Standard Model, $a_4\sim a_6$, there is a cancellation in
$A_{(\rho,\omega)K}^c$ while there isn't one in $A^c_{\pi^0 K}$. 
The suppression for $A^c_{\eta K}$ with respect to $A^c_{\eta^\prime
  K}$ has a different reason: it is due to the octet-singlet mixing,
which causes destructive (constructive) interference in the
$\eta(\eta^\prime)K$ penguin amplitude \cite{Lipkin:1980tk}. 

We stress that the numerical results presented above are often
sensitive to the approximations that we make and to the values of the
input parameters. We will refine them in the future by going beyond
the leading log approximation and taking into account uncertainties in
the input parameters other than the scale $\mu$. 
At present they should be taken as
indication to how close to $S_{\psi K_S}$ one should expect the
various $S_f$'s to be, but not as accurate predictions.
Our findings are compatible with the next to leading order 
results in the SM given by \cite{MartinCKM2005talk}.


We conclude this section by adding some general considerations
concerning the accuracy of our approximation and the stability
of our results. We discuss two main points.
First, the validity of our approximation for
branching ratios and direct CP asymmetries.
Second, the presence of large power corrections in our analysis.

Regarding the first issue, there is a substantial
effect from NLO corrections in factorization as far as the branching
ratios are concerned.
However, the impact of such NLO corrections is very moderate
for the asymmetries $S_f$, which are the main focus of our study.
This can be demonstrated explicitly in the SM by comparing our
results (Table \ref{tab:afu}) with the more complete, full-fledged NLO
analysis of \cite{MartinCKM2005talk}.
In fact, even the estimate of theoretical uncertainties, which we
obtained by a variation of the renormalization scale, is in good
agreement with \cite{MartinCKM2005talk}, 
where also other sources of uncertainty
are included. This clearly demonstrates that our approach gives a
reasonable picture for the asymmetries $S_f$, including the issue
of uncertainties and the stability against subleading effects.

In addition, our framework is still consistent with the observed
direct CP asymmetries $C_f$, which are compatible with zero.
This is the case in general for direct CP asymmetries in all
$B$ decays, where $A_{CP}(K^+\pi^-)$ is so far the only exception.
Even in this case the observed direct CP asymmetry is not very large
(about $10\%$).

Regarding the second issue, we stress again in this context 
that the dominant power corrections related
to $Q_6$ and $Q_8$ are included in our analysis (as already mentioned in
footnote 1). Further power corrections, in particular those from
annihilation topologies can be estimated \cite{Beneke:2000ry} 
to be typically of the order $10$ - $20\%$ with respect to the 
factorizable prediction.
Hence the impact of such effects is subdominant. This is also
confirmed by the comparison with \cite{MartinCKM2005talk} mentioned above.


\section{New Physics}
\label{sec:np}
\subsection{Generic analysis}
For the purpose of discussing new physics, it is convenient to define
a phase $\beta_{\rm eff}$: 
\beq\label{beff}
\lambda_{\psi K_S}\equiv-e^{-2i\beta_{\rm eff}}.
\eeq
In writing down eq.~(\ref{beff}), we are making the very plausible
assumption that the quark transition $b\to c\bar cs$ is dominated by a
single weak phase. This is clearly the case within the SM,
where the phase of $V_{cb}V_{cs}^*$ dominates to better than one
percent. But even with new physics we do not expect significant new
contributions to SM tree level processes. The approximation
of a single weak phase may be not as good as in the SM,
but it is still expected to be a very good approximation. On the other
hand, eq. (\ref{beff}) allows for new physics in $B^0-\overline{B}^0$
mixing, in which case $\beta_{\rm  eff}\neq\beta$.    

Contributions of new physics to ${\cal H}_{\rm eff}$ of
eq. (\ref{heffbone}) modify the Wilson coefficients $C_i$. (New
physics can also introduce additional operators. We do not consider
this possibility here, but our analysis can be generalized to this
case in a straightforward manner.) This modification can be parametrized 
as follows:
\beq\label{defve}
C_i(m_W)=C_i^{\rm SM}(m_W)+\ve_ie^{i\theta_i},
\eeq
where $C_i^{\rm SM}$ is the Standard Model value of the Wilson
coefficient, $\ve_i$ is real and positive, and $\theta_i$ is a phase in
the range $[0,2\pi]$.  

The modification of $A_f$ from the SM expression [eq.~(\ref{defafu})] due to
the new physics contribution [eq.~(\ref{defve})] can always be written as follows:
\beq\label{defbfu}
A_f=A_f^c\left[1+a_f^ue^{i\gamma}+\sum_i
\left(b_{fi}^c+b_{fi}^ue^{i\gamma}\right)\ve_ie^{i\theta_i}\right]. 
\eeq
In Table \ref{tab:bfu}, we give the values of the $b_{fi}^c$ parameters 
within the QCD factorization approach
\cite{Beneke:2000ry,Ali:1998eb,Beneke:2002jn}. 
The $b_{fi}^u$ for $i=3, \ldots, 10$ are related to the $b_{fi}^c$ through
 $b^u_{fi} =( |V_{ub} V_{us}^*|/|V_{cb} V_{cs}^*|)\,  b^c_{fi}$.
Note that further $b_{f 1}^c =1$ and $b_{f 1}^u=a^u_f$.

\begin{table}[p]
  \caption{The $b_{fi}^c$ ($i=1,\ldots,10$) 
    and $b_{f1,2}^u$ parameters 
    calculated in QCD factorization to leading log approximation at $\mu=m_b$
    and
    to zeroth order in $\Lambda/m_b$ (except for chirally enhanced
    corrections).
    \label{tab:bfu}}
\vspace{0.4cm}
  \centering
  \footnotesize
  \begin{tabular}{|c|c|c|c|c|c|c|}
    \hline
$f$  & $\pi^0 K$ & $\eta K$ & $\eta' K$ & $\phi K$ & $\omega K$ & $\rho^0 K$ \\ \hline
$b_{f1}^c$ & $1$     & $1$     & $1$     & $1$     & $1$    & $1$      \\
$b_{f1}^u$ & $0.052$ & $0.08$ & $0.0072$ & $0.019$ & $0.22$ & $-0.16$  \\ \hline$b_{f2}^c$ & $-0.12$ & $-0.11$ & $-0.13$ & $-0.13$ & $-0.096$ & $-0.093$ \\
$b_{f2}^u$ & $0.32$ & $0.59$ & $-0.12$ & $-0.0024$ & $1.9$ & $-1.8$ \\ \hline
$b_{f3}^c$ & $-0.34$ & $17$ & $-26$ & $-46$ & $210$ & $9.3$ \\ \hline
$b_{f4}^c$ & $-12$ & $-7.4$ & $-17$ & $-43$ & $110$ & $81$ \\ \hline
$b_{f5}^c$ & $-12$ & $-28$ & $12$ & $-43$ & $140$ & $-55$ \\ \hline
$b_{f6}^c$ & $-31$ & $-39$ & $-21$ & $-20$ & $-51$ & $-160$ \\ \hline
$b_{f7}^c$ & $-19$ & $-32$ & $3.8$ & $22$ & $80$ & $-110$ \\ \hline
$b_{f8}^c$ & $3.6$ & $-5.1$ & $20$ & $14$ & $120$ & $-1.6$ \\ \hline
$b_{f9}^c$ & $25$ & $37$ & $3.5$ & $23$ & $45$ & $-140$ \\ \hline
$b_{f10}^c$ & $12$ & $11$ & $11$ & $24$ & $-36$ & $-51$ \\ \hline
    \end{tabular}
\end{table}

With new physics, the deviation of $S_f$ of charmless modes from $S_{\psi
  K}$ depends on $b_{fi}^c$, $b_{fi}^u$, $\ve_i$ and $\theta_i$.
If enough measurements of $S_f$ and $C_f$ asymmetries become available,
one will be able to find the values of the $\ve_i$- and
$\theta_i$- parameters that account for the data. Given a small set of
measurements, one can still find $\ve_i$ and $\theta_i$ if only a small
number of operators is affected by the new physics in a significant
way. In the next section, we focus on a simple (but well motivated)
case in which the contribution of the new physics can be parametrized
in terms of a single complex parameter. 

\subsection{A single complex parameter}
\label{sec:single}
Suppose that all the $\ve_i e^{i\theta_i}$ parameters of
eq. (\ref{defve}) are related to a single complex parameter, that is,
\beq\label{vitirp}
\ve_i e^{i\theta_i}=x_i \ve e^{i\theta}\ {\rm for\ all}\ i,
\eeq
where, in a given model of new physics, the $x_i$-constants are
computable (see section \ref{sec:specific} for three specific
examples). In such cases, eq. (\ref{defbfu}) 
takes the following form:
\beq\label{sinbfc}
A_f=A_f^c\left[1+a_f^ue^{i\gamma}+
\left(b_f^c+b_f^ue^{i\gamma}\right)\ve e^{i\theta}\right]. 
\eeq
The analysis is simpler if $|b_{f}^c|\gg|b_{f}^u|$. This is the case
for all generic combinations of operators $O_i$ in eq. (\ref{curcur})
that do not involve $i=1,2$. We proceed therefore in discussing the
case where a combination of operators within the set
$i=3,\ldots,10,7\gamma,8g$ is affected by new physics. For two classes
of interesting scenarios, simple analytical expressions can be
obtained for the modification of $S_f$ and $C_f$. 

(i) The new physics contribution is dominant. Explicitly,
\beq\label{largenp}
\ve |b_{f}^c|\gg|a_{f}^u|,1.
\eeq
The shift in all modes where eq.~(\ref{largenp}) holds is
universal and depends only on $\theta$:
\beqa\label{unishi}
-\eta_f S_f&\simeq&\sin2\beta_{\rm eff}\cos2\theta+\cos2\beta_{\rm
  eff}\sin2\theta,\no\\
C_f&\simeq&0.
\eeqa

(ii) The new physics contribution and the SM pollution are
small. Explicitly,
\beq\label{smallnpau}
\ve |b_{f}^c|\ll1\ {\rm and}\ |a_{f}^u|\ll1.
\eeq
The shift is mode dependent and depends on $\ve\sin\theta$:
\beqa\label{smallnp}
-\eta_f S_f&\simeq&\sin2\beta_{\rm eff}+2\cos2\beta_{\rm eff}\left[{\cal
  R}e(b_{f}^c)\ve\sin\theta+{\cal
  R}e(a_f^u)\sin\gamma\right],\no\\ 
C_f&\simeq&-2{\cal I}m(b_{f}^c)\ve\sin\theta-2{\cal
  I}m(a_f^u)\sin\gamma. 
\eeqa

Since the $b_{f}^c$ values are known (see Table \ref{tab:bfu}), the
pattern of deviations is predicted and can be used, by comparing with
the data, to close in on the set of operators that is responsible for
the deviations and to extract $\ve\sin\theta$.

In case that none of these approximations applies, one can still find 
the pattern of deviations by numerical evaluation for any values for
$\ve$ and $\theta$. By comparing these theoretical $\delta S_f$
and $C_f$ to the data, allowed regions in the $\ve-\theta$ plane
(for each scenario) can be determined, and a best fit point (among all
scenarios) can be found. We demonstrate this procedure in the next section
by using present data.

\subsection{Specific Scenarios}
\label{sec:specific}
We now consider three explicit examples of new physics models. All
three examples are well motivated and fulfill eq.~(\ref{vitirp}).

(i) The dominant effect of the new physics is through $Z$-penguins
\cite{Atwood:2003tg}. Then, the Wilson coefficients are modified as
follows: 
\begin{eqnarray}\label{zpeng}
C_3^{\rm NP}(m_W)&=&\frac16 \ve_z e^{i\theta_z},\nonumber\\
C_7^{\rm NP}(m_W)&=&\frac23 \sin^2\theta_W \ve_z e^{i\theta_z},\nonumber\\
C_9^{\rm NP}(m_W)&=&-\frac23(1-\sin^2\theta_W)\ve_z e^{i\theta_z}.
\end{eqnarray}
Here
\begin{equation}
\ve_z e^{i\theta_z}= \frac{g^2}{4\pi^2}\frac{Z_{sb}}{V_{tb}V_{ts}^*},
\end{equation}
where $Z_{sb}$ parametrizes the coupling of the $Z^0$ boson to the left
handed strange and bottom quarks. The consistency of the measured
BR$(b\to s\ell^+\ell^-)$ with the SM prediction requires that
$|Z_{sb}|\leq0.08$ \cite{Atwood:2003tg} which gives 
\begin{equation}
\ve_z\lsim 0.02.
\end{equation}

(ii) Kaluza-Klein excitations of gluons in RS models
\cite{Burdman:2003nt} couple mainly to the third generation quarks and
contribute to all gluonic penguin operators:\footnote{In
  ref. \cite{Agashe:2004ay} it is argued that, quite generically in 
the RS framework, the set of operators in eq. (\ref{kakl}) is
subdominant and constrained to give only a small contribution to the
decays in question. However, this issue is model dependent, and eq.~(\ref{kakl})  also
represents a viable scenario.}
\begin{eqnarray}\label{kakl}
C_4^{\rm NP}(m_W)&=&C_6^{\rm NP}(m_W)=\ve_k e^{i\theta_k},\nonumber\\
C_3^{\rm NP}(m_W)&=&C_5^{\rm NP}(m_W)=-\frac13 \ve_k e^{i\theta_k}.
\end{eqnarray}
Here 
\begin{equation}
\ve_ke^{i\theta_k}= -\pi\alpha_s(M_G)\left(\frac{v}{M_G}\right)^2
\frac{D_{sb}}{V_{tb}V_{ts}^*}\chi,
\end{equation}
where $D_{sb}$ is the rotation matrix from interaction to mass basis
for the left-handed down quarks, $\chi={\cal O}(1)$ is a model
dependent parameter (related to bulk fermion masses), and $M_G$ is the
mass of the lightest excitation. With $M_G\gsim1\ TeV$ and $|D_{sb}|\sim
|V_{tb}V_{ts}^*|$, we expect
\beq
\ve_k\lsim 0.02.
\eeq

(iii) Enhanced chromomagnetic operator \cite{Kagan:1997sg} can be
parametrized as follows: 
\begin{equation}\label{chma}
C_{8g}^{\rm NP}(m_W)=\ve_{g}e^{i\theta_g}.
\end{equation}
(We still make the approximation that $C_{8g}^{\rm SM}(m_W)=0$ and
neglect further $C_{8g}^{\rm SM}(m_b)$ induced by $O_{1}$.)
From BR($B \to X_s g) <0.09$ \cite{Coan:1997ye,Kagan:1997sg}, we have  
\beq
\ve_g\lsim1.
\eeq
Note that in this case the new physics contribution enters at next to
leading order. Yet, the modifications of the CP asymmetries can be
substantial. An experimental improvement by a factor of a few in
the upper bound on BR($B \to X_s g)$ will restrict the potential
modifications in this scenario in a significant way. 

We have calculated the $b_f^c$ and $b_f^u$ parameters of
eq. (\ref{sinbfc}) for each of the three scenarios. We use the
factorized decay amplitudes from
\cite{Beneke:2000ry,Ali:1998eb,Beneke:2002jn}. (We use asymptotic distribution amplitudes in the analysis of the scenario with an enhanced chromomagnetic operator.) 
The results are presented in Table \ref{tab:sce}. The $b_f^c$ parameters are quite stable under variation of the scale. As anticipated in section
\ref{sec:single}, in all three scenarios we 
have $b^u_f = (|V_{ub} V_{us}^*|/|V_{cb} V_{cs}^*|) \,  b^c_f$.

\begin{table}[p]
  \caption{The $b_{f}^c$ parameters calculated in QCD
    factorization to leading log approximation and to zeroth order in
    $\Lambda/m_b$ (chirally enhanced corrections are
    included) in the three scenarios: (i) $Z$-penguins [eq.
    (\ref{zpeng})], denoted with subindex $z$; (ii) KK-gluons
    [eq. (\ref{kakl})], denoted with subindex $k$; (iii)
    Chromomagnetic operator [eq. (\ref{chma})], denoted with
    subindex $g$.  The first row in each case is obtained for $\mu=m_b$,
whereas the second row gives the values for $\mu=m_b/2$ (first) and
 $2 m_b$ (second).
    \label{tab:sce}}
\vspace{0.4cm}
  \centering
  \footnotesize
  \begin{tabular}{|c|c|c|c|c|c|c|}
    \hline
$f$  & $\pi^0 K$ & $\eta K$ & $\eta' K$ & $\phi K$ & $\omega K$ & $\rho^0 K$ \\ \hline
$b_{fz}^c$ & $-16$ & $-21$ & $-5.4$ & $-16$ & $24$ & $57$      \\
 & $\left[ -12,-22 \right]$ & $\left[-16,-30 \right]$ & $\left[-4.0, -7.6 \right]$ & $\left[ -10,-24 \right]$ & $\left[ 29,28\right]$ & $\left[60,65 \right]$   \\ \hline
$b_{fk}^c$ & $-39$ & $-43$ & $-34$ & $-33$ & $-61$ & $-64$ \\
 & $\left[ -28,-55 \right]$ & $\left[-33,-59 \right]$ & $\left[ -22,-51 \right]$ & $\left[-24,-48 \right]$ & $\left[-63,-74 \right]$ & $\left[-66,-75 \right]$   \\ \hline
$b_{fg}^c$ & $0.75$ & $0.64$ & $0.86$ & $1.4$ & $-1.7$ & $-1.5$ \\
 & $\left[0.61, 0.99 \right]$ & $\left[0.48, 0.87 \right]$ & 
$\left[0.76, 1.07 \right]$ & $\left[1.1, 1.8 \right]$ & $\left[-2.8, -1.4 \right]$ & 
$\left[-2.0 ,-1.3 \right]$ \\ \hline
    \end{tabular}
\end{table}

Given the upper bounds on $\ve$ for the three scenarios --
$\ve_z\lsim0.02$, $\ve_k\lsim0.02$ and $\ve_g\lsim1$ -- we learn that in
none of the three examples can the new physics strongly dominate over
the SM contributions, that is, $\ve b_f^c\not\gg a_f^u,1$. Thus,
if the data makes a convincing case for a universal pattern of
deviations, it would mean that either this pattern is accidental or
that the new physics is different from the three specific scenarios
discussed here.

On the other hand, the interpretation of the results presented in
Table \ref{tab:sce} is straightforward in the other limit
discussed in the previous subsection. For the four modes $f=\pi^0
K_S,\ \eta K_S,\ \eta^\prime K_S$ and $\phi K_S$, we have, say,
$|b_f^c\ve | \lsim0.3$ for $\ve_z\lsim0.01$, $\ve_k\lsim0.005$, and 
$\ve_g\lsim0.17$. In any of these cases, the deviation from the SM
prediction is proportional to $b_f^c$ [see
eq. (\ref{smallnp})]. Consequently, there is a distinct pattern of
deviations for each of the three scenarios. For example, the deviation
of $S_{\eta^\prime K_S}$ from the SM prediction is a factor of four 
smaller than that of $S_{\eta K_S}$ in the first scenario
($\ve_z\neq0$), a factor $\sim1.3$  smaller in the second
($\ve_k\neq0$) and a factor of $\sim1.3$  larger in the third
($\ve_g\neq0$). The deviation of $S_{\phi K_S}$ is similar or somewhat smaller
than that of $S_{\pi K_S}$ in the first two scenarios, but a factor of
$\sim1.9$ larger in the third. 
The second scenario has the characteristic feature that the SM deviations
of $S_{\phi K_S}$ and $S_{\eta^\prime K_S}$ are very similar, whereas in scenario i (iii)  $S_{\phi K_S}$ is a factor of 3 (1.6) larger than 
$S_{\eta^\prime K_S}$.
 It is interesting to note that in
all three new physics examples, the deviations for $b_f^c \ve\ll1$ in the 
modes that have been most accurately measured, $f=\pi^0
K_S,\ \eta^\prime K_S$ and $\phi K_S$, and also $\eta K_S$, are in the
same direction, that is, the four $S_f$ asymmetries are either all
larger or all smaller [depending on sign($\sin\theta$)] than the SM
prediction.

\begin{figure}
  \centering
  \parbox[b]{0.5\textwidth}{\includegraphics[width=\linewidth]{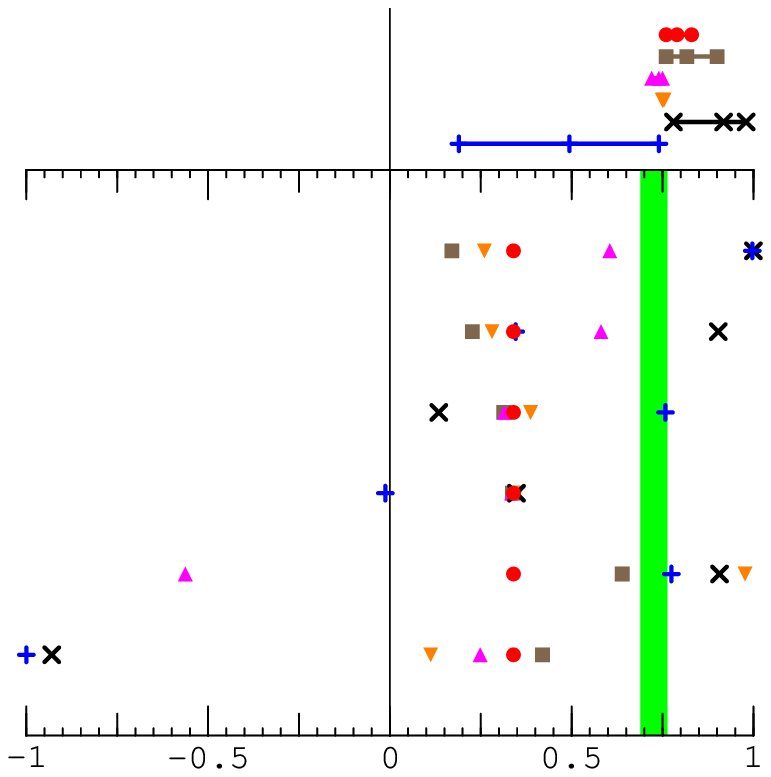}}~$S_f$
  \caption{Pattern of $S_f$ asymmetries for the three scenarios
    presented in section \ref{sec:specific} with fixed $S_{\pi
      K_S}=0.34$. The symbols represent ${\color[rgb]{1,0,0} \bullet}
    = \pi$, ${\color[rgb]{0.5,0.4,0.3} \blacksquare} = \eta$,
    ${\color[rgb]{1,0,1}\blacktriangle} = \eta'$,
    ${\color[rgb]{1,0.5,0}\blacktriangledown} = \phi$,
    ${\color[rgb]{0,0,0}\boldsymbol{\times}} = \omega$ and
    ${\color[rgb]{0,0,1}\boldsymbol{+}} = \rho$.  The upmost pattern
    gives the asymmetries in the SM for $\mu=m_b/2,m_b$ and $2 m_b$.
    The other six patterns, from top to bottom, correspond to
    $(\ve_z,\theta_z)=(0.02,0.85)$, $(\ve_z,\theta_z)=(0.02,1.79)$,
    $(\ve_k,\theta_k)=(0.02,0.12)$, $(\ve_k,\theta_k)=(0.02,2.52)$,
    $(\ve_g,\theta_g)=(1,3.27)$, and $(\ve_g,\theta_g)=(1,5.64)$. The
    vertical bar represents the experimental value for $S_{\psi
      K_S}$.}
  \label{fig:S}
\end{figure}

To demonstrate how the pattern of deviations probes new physics, we
perform the following exercise. For each of the three scenarios
discussed in this section, we take $\ve_i$ at the phenomenological
upper bound, that is $\ve_z=0.02$, $\ve_k=0.02$ and $\ve_g=1$. Since
the new physics contribution is now as large as it can get, the
patterns of deviations are expected to be more significant. We then
pick the two values of $\theta_i$ for which $S_{\pi K_S}=0.34$, the
experimental central value. (There is no particular reason for
selecting $S_{\pi K_S}$. Our aim is just to demonstrate the idea that
the pattern of deviations is sensitive to new physics.) We present the
results in Fig. \ref{fig:S}. One can clearly see that the six patterns
are different from each other, so that each scenario can in principle
be easily tested.

\section{Present Data}
\label{sec:pre}
At present, the B factories provide constraints on eight relevant CP 
asymmetries. The data are presented in Table \ref{tab:fcp}. We would
like to test various scenarios of new physics by performing a $\chi^2$
fit to the data. Since we are mainly interested in demonstrating the
potential power of the data to probe new physics in the future, we
make several simplifications:
\begin{enumerate}
\item We work within the framework defined in
  section \ref{sec:np}.B, that is, new physics contributions that
  depend on only a single complex parameter.
\item Since the measurement of $S_{\psi K_S}$ is very accurate, we fix
  the value of $\sin2\beta_{\rm eff}$ to its central value, and do not include
  it as a fit parameter. Furthermore, since experimental data
  \cite{Aubert:2004cp} disfavor $\cos2\beta_{\rm eff}<0$, we allow
  only positive $\cos2\beta_{\rm eff}$ values.
\end{enumerate}
We thus calculate, for various scenarios,
\beq
\chi^2(\ve,\theta)\equiv \sum_{f}\left[\frac{X_f^{\rm exp}-X_f^{\rm
    QF}(\ve,\theta)}{\sigma_f^{\rm exp}}\right]^2,
\eeq
where $X_f^{\rm exp}$ and $\sigma_f^{\rm exp}$ are, respectively, the
central value and the one sigma range of the experimental
measurements of $S_f$ and $C_f$ for $f=\pi^0K_S$, $\eta^\prime K_S$,
$\phi K_S$ and $\omega K_S$, and $X_f^{\rm QF}(\ve,\theta)$ are the
(central) theoretical values of $S_f$ and $C_f$ calculated within the 
QCD factorization
approach for given values of $\ve$ and $\theta$. Only experimental 
uncertainties are taken into account. Since there are
eight observables and two free parameters, $\chi^2$ has a six degrees
of freedom distribution.

We consider eleven cases. Firstly, we investigate the possibility that
the new physics contributes significantly via only a single $O_i$,
where $i=3,\ldots,10$. (Specific models of new physics do not provide
a strong motivation for such scenarios. Yet, this investigation is
useful in the two limits discussed in section \ref{sec:single}. If the
new physics is dominant and depends on a single weak phase, we can see
the universality of its effects [eq.~(\ref{unishi})]. If the new
physics contribution is small, the effects are well approximated by a
sum over the shifts from the various 
operators [eq.~(\ref{smallnp})].) In each case, we perform our
analysis for $\ve\leq5$ and all $\theta\in[0,2\pi]$. Secondly, we
investigate the three specific examples presented in section
\ref{sec:specific}. We do this for $\ve$ within the phenomenologically
allowed ranges and for all $\theta$ values. 

\begin{figure}[htbp]
  \centering
  \subfigure[]{
    \begin{minipage}[c]{0.30\textwidth}
      $\ve_4$ \parbox[c]{0.8\textwidth}{\includegraphics[width=0.8\textwidth]{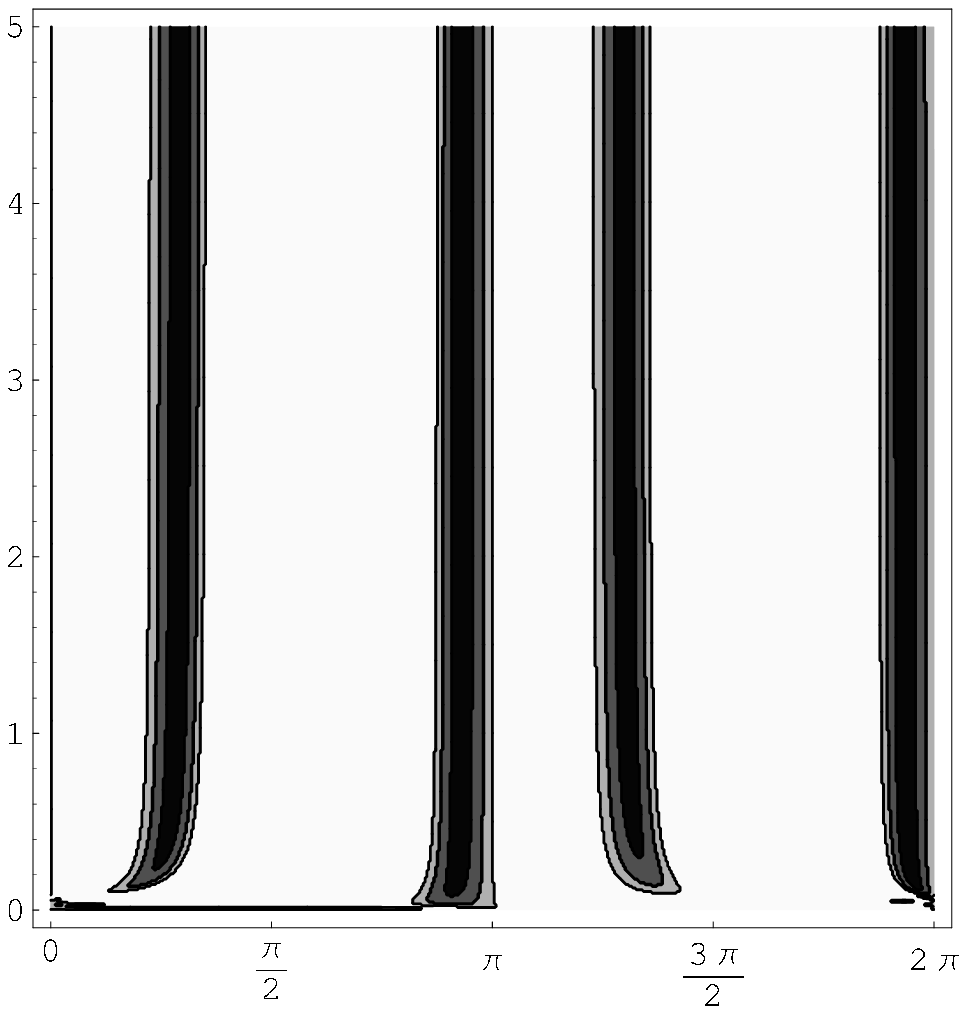}}\\
      $\theta_4$ 
    \end{minipage}
  }
  \subfigure[]{
    \begin{minipage}[c]{0.30\textwidth}
      $\ve_5$ \parbox[c]{0.8\textwidth}{\includegraphics[width=0.8\textwidth]{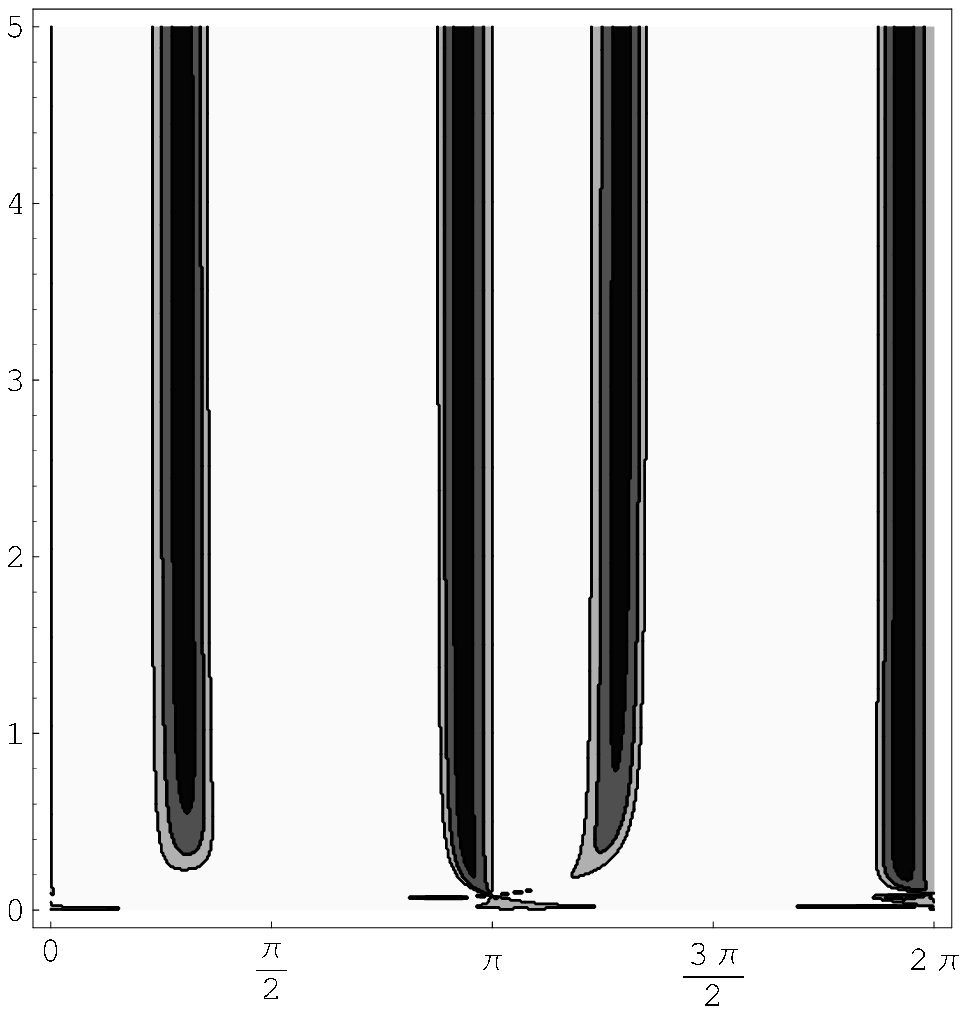}}\\
      $\theta_5$ 
    \end{minipage}
  }
  \subfigure[]{
    \begin{minipage}[c]{0.30\textwidth}
      $\ve_9$ \parbox[c]{0.8\textwidth}{\includegraphics[width=0.8\textwidth]{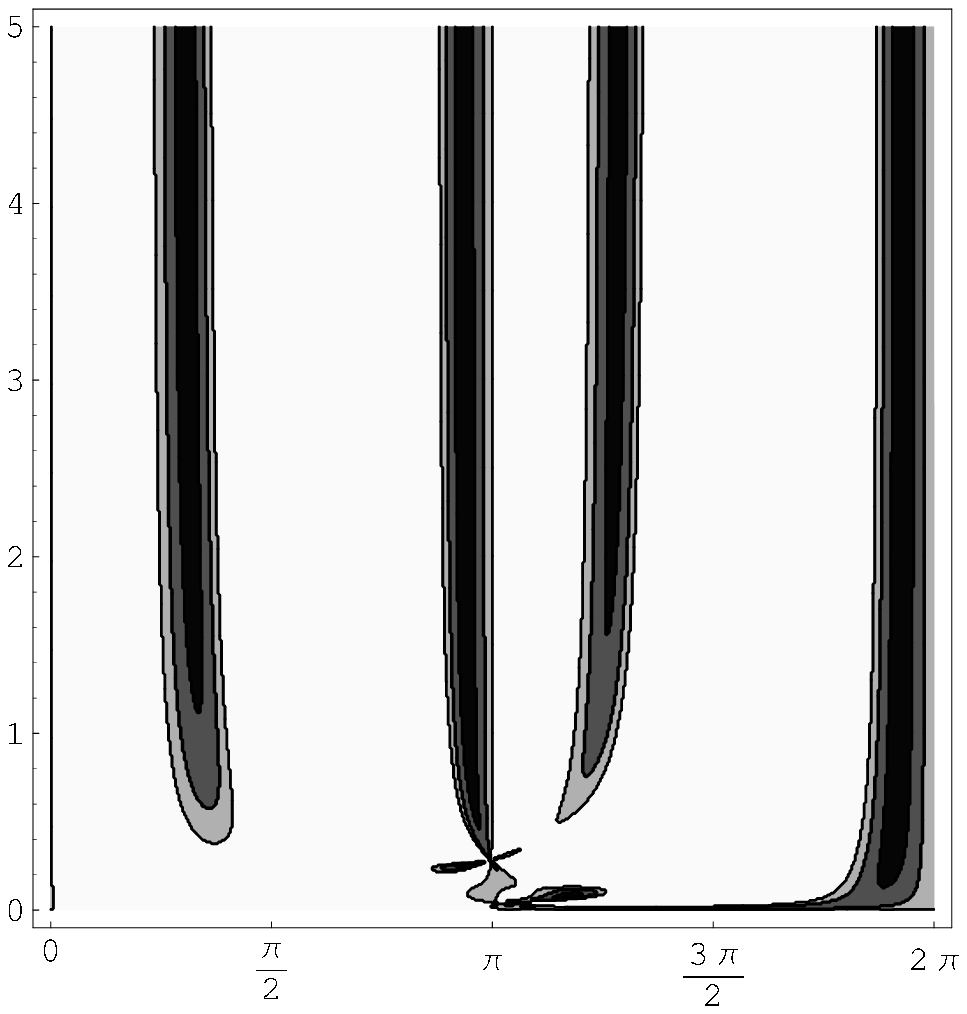}}\\
      $\theta_9$ 
    \end{minipage}
  }
  \caption{Allowed region in the $\ve_i-\theta_i$ plane for new
    physics in a single operator: (a) $O_4$, (b) $O_5$ and (c) $O_9$.
    The black, dark grey, and light grey regions correspond to
    probability higher than 0.32, 0.046, and 0.0027, respectively.}
  \label{fig:individual}
\end{figure}

\begin{figure}[htbp]
  \centering
  \subfigure[]{
    \label{fig:Zpenguin}
    \begin{minipage}[c]{0.30\textwidth}
      $\ve_z$ \parbox[c]{0.8\textwidth}{\includegraphics[width=0.8\textwidth]{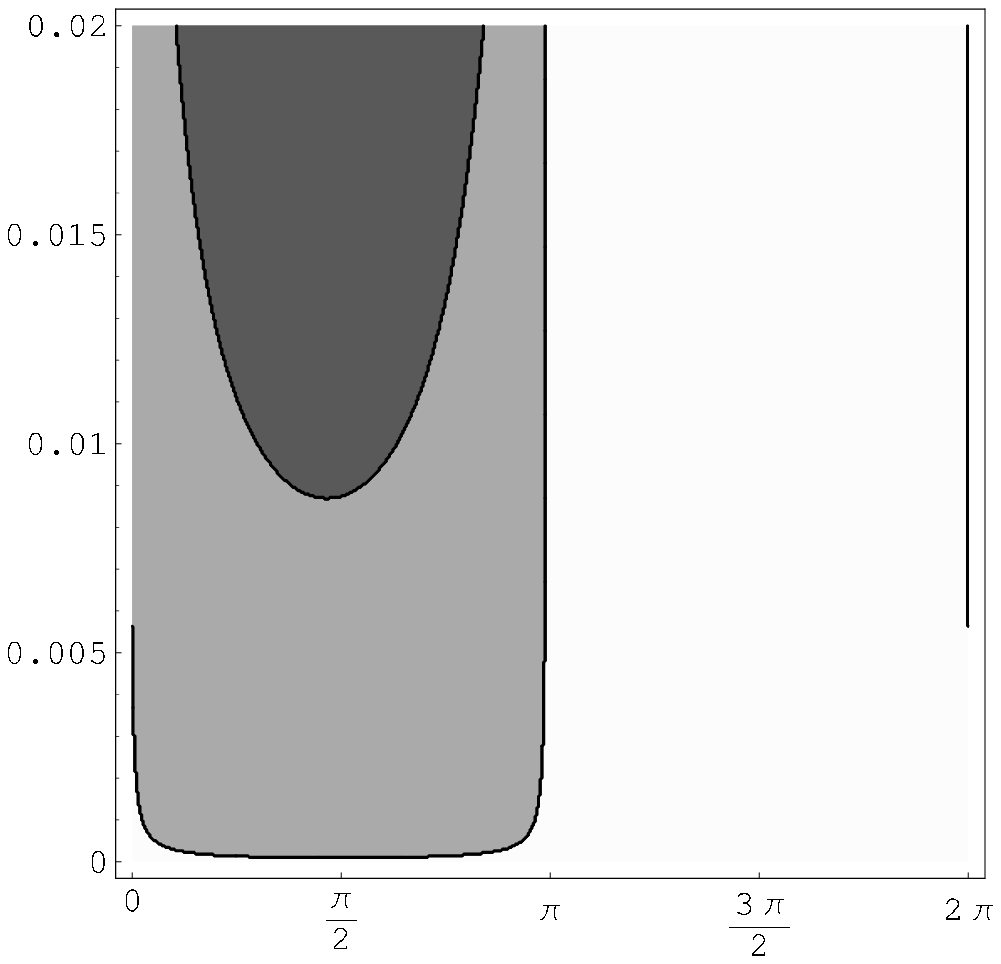}}\\
      $\theta_z$ 
    \end{minipage}
  }
  \subfigure[]{
    \label{fig:KKgluon}
    \begin{minipage}[c]{0.30\textwidth}
      $\ve_k$ \parbox[c]{0.8\textwidth}{\includegraphics[width=0.8\textwidth]{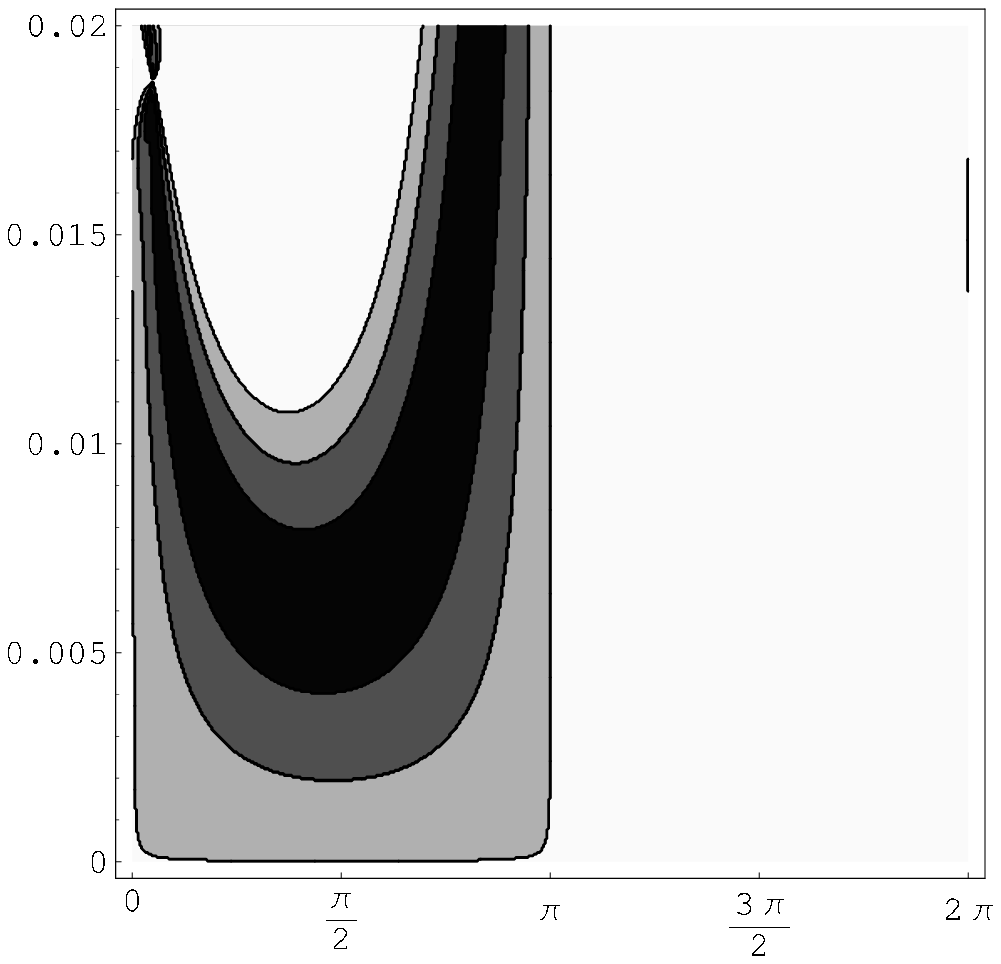}}\\
      $\theta_k$ 
    \end{minipage}
  }
  \subfigure[]{
    \label{fig:C7}
    \begin{minipage}[c]{0.30\textwidth}
      $\ve_g$ \parbox[c]{0.8\textwidth}{\includegraphics[width=0.8\textwidth]{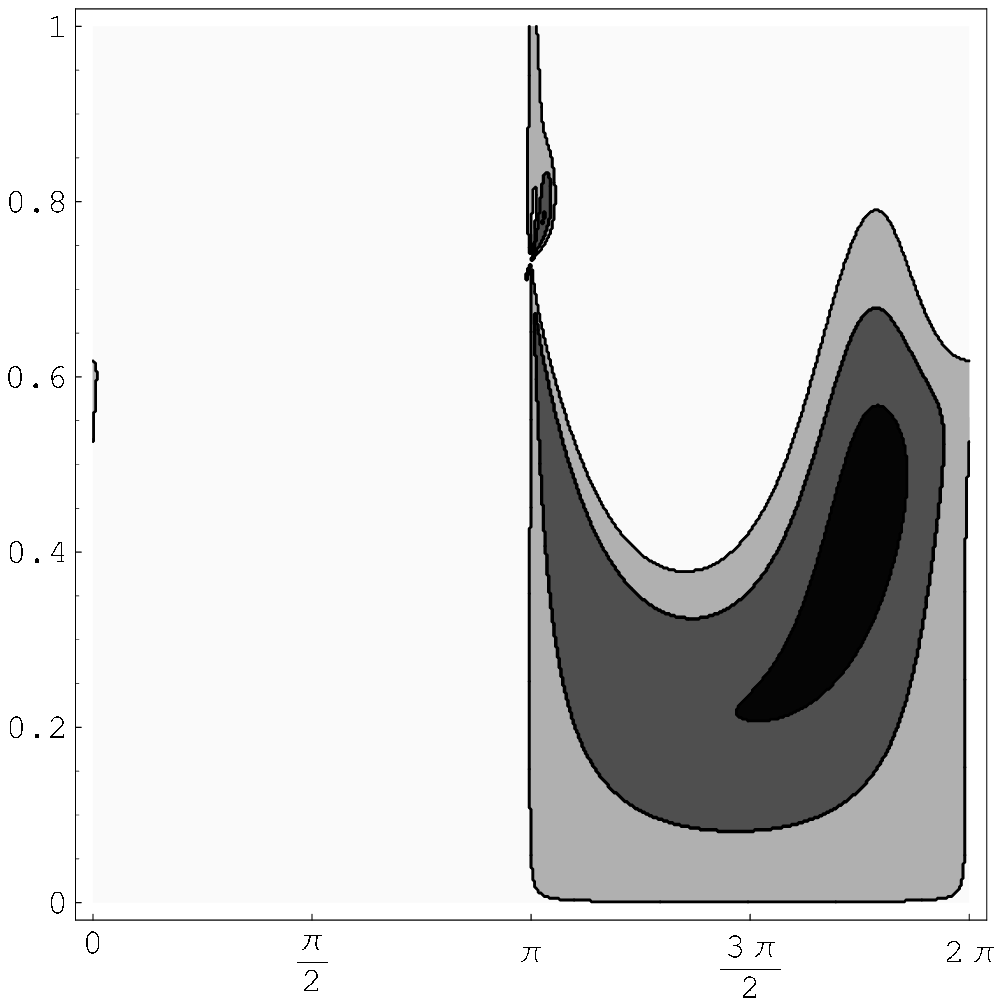}}\\
      $\theta_g$ 
    \end{minipage}
  }
  \caption{Allowed regions in the $\ve-\theta$ plane for the three
    specific scenarios, (a) enhanced $Z$-penguins, (b) KK gluons in RS
    models, (c) enhanced chromomagnetic operator. The black, dark
      grey, and light grey regions correspond to probability higher
      than 0.32, 0.046, and 0.0027, respectively.}
  \label{fig:scenario}
\end{figure}

We use the $\chi^2$ fit in two ways. For each case we find the favored
regions in the $\ve-\theta$ plane. The results are shown in
Figs.~\ref{fig:individual} and~\ref{fig:scenario}.  Since the fits to
individual operators $O_{3 \ldots 10}$ give very similar results for
large $\epsilon$, we only show the ones for $O_4, O_5$ and $O_9$.
Then we test the overall goodness of the fit for each case and learn
whether the possibility that the data are accounted for by a new
physics contribution to this (set of) operator(s) is disfavored. We
draw the following conclusions:
\begin{enumerate}
\item For almost all the Wilson coefficients, a very good fit is
obtained for a high $\ve$ value, such that eq.~(\ref{largenp}) holds and
the shift is universal [eq.~(\ref{unishi})]. The reason is that the central
values of the three best measured $S_f$ asymmetries ($f=\eta^\prime
K_S,\phi K_S,\pi^0 K_S$) show an approximately universal pattern:
$\delta S_f\sim-0.4$. 
\item For operators where eq.~(\ref{largenp}) gives the best fit, the
preferred value of $\theta$ corresponds to
$\sin2(\beta+\theta)\sim0.4$. Note that a four-fold ambiguity arises,
giving $\theta_i\sim55^o,\ 170^o,\ 235^o,\ 350^o$, which can be seen
in Fig.~\ref{fig:individual}. Had we allowed for
$\cos2\beta_{\rm eff}<0$, there would be eight possible solutions for
high $\ve$. 
\item For each of the three specific scenarios that we considered,
there exists a region in the $\ve-\theta$ plane where a good fit is
obtained. Note that within the operator basis we use, 
in the first two scenarios with modified four-Fermi operators a phase 
smaller than $\pi$ is preferred, 
whereas $\pi \leq \theta_g <2 \pi$ is favored.
\item Naively, the probability that there is no new physics
contribution affecting the measured $S_f$'s  is small. (The SM point,
$\ve=0$, lies in the white area.) One has to bear
in mind, however, that we used a rather crude approximation, and that
there are large uncertainties in our evaluation of the matrix
elements. Furthermore, the discrepancy between $S_{\eta^\prime
 K_S}^{\rm exp}=0.43\pm0.11$ and $S_{\eta^\prime
 K_S}^{\rm QF}=0.74$ plays a significant role in this
result. Note however that the Belle and Babar results for this
observable are not quite consistent. Inflating the error
according to the PDG prescription would yield $S_{\eta^\prime
 K_S}^{\rm exp}=0.43\pm0.17$ and a better SM fit. (Now the SM point
lies in the light grey area.)
\end{enumerate}

\section{Conclusions}
\label{sec:con}
$B$-meson decays that proceed via the quark transition $b\to q\bar q
s$, for $q=s,d,u$, open a window to new physics because the SM
contribution is either loop- or CKM-suppressed. In particular, the CP
violating asymmetries $S_f$ and $C_f$ in these modes may reveal the
existence of new sources of CP violation beyond the KM phase.

With new physics, these decays may get significant contributions that
depend on a phase that is different from the SM phase. Then, not only
will these asymmetries be different from the SM prediction but also,
in general, they will differ from each other. The pattern of
deviations, $\delta S_f=-\eta_f S_f-S_{\psi K}$ and $C_f$,
allows us to probe in detail the nature of the new physics that
accounts for the effect.

Factorization schemes predict that there is an accidental cancellation
between the leading Standard Model contributions to the $\eta K_S$,
$\omega K_S$ and $\rho K_S$ modes (see section \ref{sec:for}). The
suppression of the Standard Model amplitudes makes the corresponding
CP asymmetries more sensitive to New Physics. At the same time, the
sensitivity to hadronic uncertainties is also stronger. 

We presented a way to analyze the data in a (new physics) model
independent way. In a low energy effective theory, the effects of the
new physics are expressed as modifications of Wilson coefficients in
the $\Delta B=1$ effective Hamiltonian. The data can be used to find
which Wilson coefficients are modified and by what (complex) factor.

The analysis is simplified when the modification of the Wilson
coefficients depends on a single complex parameter, as is often the
case in specific models. We gave analytic
expressions for the modifications in some interesting limits. In
particular, for $C_{3,\ldots,10,8g}$, when the new physics
dominates, a universal shift of the asymmetries arises [see
eq. (\ref{unishi})]. If the new physics contribution is small, then
the deviations (for $f=\eta^\prime K_S,\phi K_S,\pi^0 K_S,\eta K_S$)
have a pattern that depends in a simple way on the $b_f^c$-parameters
[see eq. (\ref{smallnp})].

We applied this method to present data, using lowest order QCD
factorization to calculate the hadronic matrix elements. We find that,
since the best three measured $S_f$'s have similar central values
($\sim0.4$) which are, however, different from $S_{\psi K}$, and all 
measured $C_f$'s are consistent with zero, a good fit can be
achieved for almost all operators if they are enhanced by the new
physics to a level where they dominate over the SM contribution. 
In three specific scenarios that we considered, such new physics
dominance cannot be realized in Nature because of phenomenological
constraints. Still, a good fit can be obtained for each of these
scenarios.

Our analysis can be improved in several ways. In particular, the
calculation should go beyond the leading log approximation and
incorporate power corrections. The combination of
improved calculations, more accurate measurements and additional modes
explored experimentally, is likely to make the pattern of deviations
from the SM (or their absence) a powerful probe of flavor and
CP violating new physics.

\acknowledgments We thank Martin Beneke for communications and Zoltan
Ligeti and Lincoln Wolfenstein for comments on the manuscript. This
work was supported by a grant from the G.I.F., the German--Israeli
Foundation for Scientific Research and Development.  Y.N.~is supported
by the Israel Science Foundation founded by the Israel Academy of
Sciences and Humanities, by EEC RTN contract HPRN-CT-00292-2002, by
the Minerva Foundation (M\"unchen), by the United States-Israel
Binational Science Foundation (BSF), Jerusalem, Israel and by
Fundaci\'on Antorchas/Weizmann.


\end{document}